# Terahertz hyperspectral imaging with dual chip-scale combs

Lukasz A. Sterczewski[1,2,†], Jonas Westberg[1,†], Yang Yang[3], David Burghoff[3,4], John Reno[5], Qing Hu[3,*], and Gerard Wysocki[1,**]

[1]Department of Electrical Engineering, Princeton University, Princeton NJ 08544, USA
[2]Faculty of Electronics, Wroclaw University of Science and Technology, Wroclaw 50370, Poland
[3]Department of Electrical Engineering and Computer Science, Research Laboratory of Electronics, Massachusetts Institute of Technology, Cambridge, MA 02139, USA
[4]Department of Electrical Engineering, University of Notre Dame, Notre Dame, IN 46556, USA
[5]Center for Integrated Nanotechnology, Sandia National Laboratories, Albuquerque, NM 87123, USA

*qhu@mit.edu
**gwysocki@princeton.edu
†These authors contributed equally to this work.

**Hyperspectral imaging is a technique that allows for the creation of multi-color images. At terahertz wavelengths, it has emerged as a prominent tool for a number of applications, ranging from non-ionizing cancer diagnosis [1,2] and pharmaceutical characterization [3,4] to non-destructive artifact testing [5,6]. Contemporary terahertz imaging systems typically rely on non-linear optical down-conversion of a fiber-based near-infrared femtosecond laser, requiring complex optical systems. Here, we demonstrate hyperspectral imaging with chip-scale frequency combs based on terahertz quantum cascade lasers. The dual combs are free-running and emit coherent terahertz radiation that covers a bandwidth of 220 GHz at 3.4 THz with ~10 μW per line. The combination of the fast acquisition rate of dual-comb spectroscopy with the monolithic design, scalability, and chip-scale size of the combs is highly appealing for future imaging applications in biomedicine and in the pharmaceutical industry.**

Optical imaging techniques have long been indispensable in the natural sciences, with widespread adoption in the fields of biology, astronomy, material science and medicine. If merged with broadband spectroscopy, the traditional two-dimensional image is expanded into its hyperspectral counterpart, where each pixel represents a spectrum. This transforms the image into a three-dimensional hyperspectral data cube. The initial interest in this field was predominantly driven by defense and astronomical applications, but the rapid advances in electronics and computing during recent decades have precipitated widespread adoption of hyperspectral imaging in diverse scientific areas, such as biomedicine, agriculture, and environmental sensing.

Terahertz (THz) technologies have also experienced a similar growth over the past decades [7–10], propelled in large part by the significant advances in ultrafast lasers and non-linear optics that have transferred THz generation from bulky gas lasers [11] to more user-friendly fiber-based femtosecond laser systems [12]. For non-invasive imaging [13,14], THz radiation exhibits a number of attractive properties, most notably the ability to propagate through materials that are opaque at higher optical frequencies, including fabrics, papers and plastics. This property has driven developments in non-destructive quality control and security imaging to awareness beyond the scientific community. For spectroscopic applications, the excitation of vibrational modes at THz frequencies in many inorganic and organic compounds is highly compelling, not only for chemical identification, but also for discrimination based on the crystal lattice arrangement. In addition, the high sensitivity to water absorption together with the non-ionizing property of THz radiation makes this part of the electromagnetic spectrum highly suitable for bio-imaging applications, e.g. identification of different types of human tissue based their water content [15]. Another area of active THz imaging development is the pharmaceutical field, where THz radiation can be used to characterize tablet coating thicknesses [3,4], tablet composition, and degradation of the active ingredients [16].

To date, the most commonly used THz spectroscopy systems are based on interferometric or on time-domain sampling techniques, which enable wide spectral coverage with good sensitivities and robust operation, but with significant fundamental limitations. The interferometric Fourier Transform spectroscopy (FTS) technique [17], originally developed for the visible and infrared spectral regions, offers unrivaled spectral coverage and ease-of-use. Unfortunately, THz-FTS suffers from large footprint, low brightness and slow acquisition, as a consequence of using a black-body radiation source and an opto-mechanically scanned mirror. Nevertheless, the broad optical bandwidth and proven reliability of THz-FTS ensures its use in sectors of far-infrared spectroscopy and imaging where ample averaging is permitted.

THz time-domain spectroscopy (THz-TDS) usually relies on electro-optical or photoconductive sampling to reconstruct the electromagnetic field of a THz pulse, where the short pulse durations of modern mode-locked fiber lasers make it inherently broadband with good signal-to-noise ratio (SNR) in the 1-3 THz range. However, just as in FTS systems, the time-domain sampling typically relies on a mechanically scanned optical delay line, which limits the acquisition speed and enlarges the system footprint.

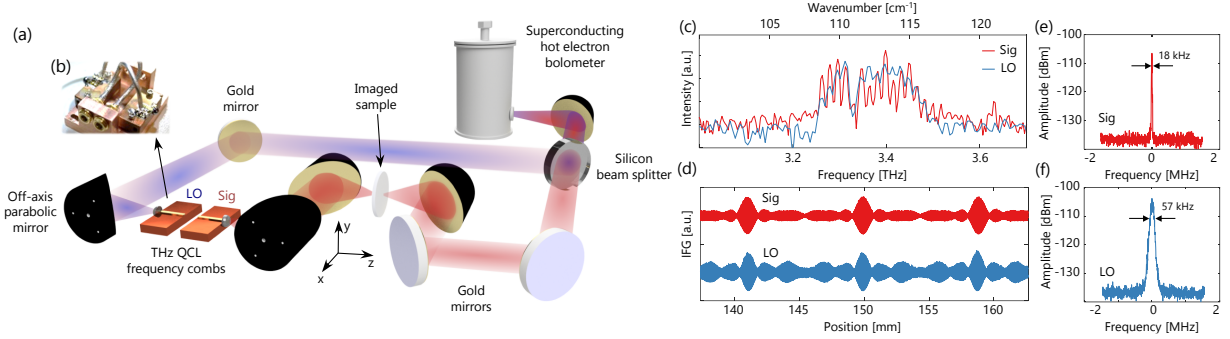

**Fig. 1.** (a) Two THz-QCL OFCs are aligned antiparallel and their outputs are collected by two off axis parabolic gold-coated mirrors. The signal OFC is focused on a sample (solid disc) placed on a XYZ-translation stage. The transmitted light is collected and combined with the local oscillator comb on a hot electron bolometer with a bandwidth of 5 GHz. (b) Photo of the two chip-scale THz-QCLs mounted on a copper submount. (c) Optical mode spectra for the two THz combs centered at around 3.4 THz. (d) Interferograms (IFGs) corresponding to the spectra shown in (c), measured with a THz Fourier transform spectrometer. (e), (f) Intermode beat notes from the two THz-QCL OFCs. The center frequencies are 16.988 GHz and 17.027 GHz for (e) and (f), respectively.

Over the last two decades, a novel approach to reconstruct the optical spectrum of a repetitive pulse train has increasingly gained popularity, the dual-comb spectroscopy (DCS) technique [18]. In DCS, a second pulse generator with different repetition rate is used to asynchronously sample the pulse train of the first, a process that results in a time-domain signal in the form of an interferogram, analogous to that obtained in FTS. This scheme eliminates the need for any opto-mechanical movement, which enables high acquisition speeds (μs) and user-friendly operation. Although most of this field primarily involves commercially available fiber-laser based optical frequency combs (OFCs) operating in the near-infrared, its usefulness has also been demonstrated in both the mid-infrared [19,20] and the THz [21] using nonlinear media for frequency conversion. In 2012, a fundamentally different OFC was demonstrated in the mid-infrared, the quantum cascade laser (QCL) OFC [22], which exploits the nonlinearity of a low-dispersion, electrically pumped, semiconductor gain medium to directly emit comb radiation around an optical frequency defined via careful control of the layered semiconductor growth. This first demonstration was shortly followed by an extension to the THz-domain [23], where the intrinsically large device dispersion was compensated by waveguide engineering. Much of the strength of the QCL-OFCs lies in their high optical power per mode, which can reach several mW in the mid-infrared [24] and tens of μW in the THz [25] making spectroscopic assessments of highly absorbing media possible. This is especially advantageous when considering spectroscopy of liquids or solids where a trade-off between the GHz-range frequency resolution and the high optical power per mode can be justified. Several examples of chip-scale semiconductor laser-based DCS have been demonstrated in the mid-infrared [26–29] and THz [25] with peak SNRs of more than $10^4/\sqrt{s}$ [27,29].

Our proof-of-concept THz hyperspectral imaging system is based on a pair of dispersion compensated THz-QCL OFCs employed in the asymmetric dual-comb configuration. The system achieves a spectral coverage of ~220 GHz at a center wavelength of 3.4 THz with more than 100 μW of optical power emitted by each comb. A hyperspectral composite image of a pressed disc consisting of α-D-glucose monohydrate (GMH), α-D-lactose monohydrate (LMH), and L-histidine monohydrochloride monohydrate (LHHM) is obtained via a raster scan of the sample. This demonstration of hyperspectral imaging using electrically-pumped semiconductor laser frequency combs, fabricated using scalable technology, opens possibilities of future compact hyperspectral imaging systems for applications in biomedicine, biochemistry and the pharmaceutical industry.

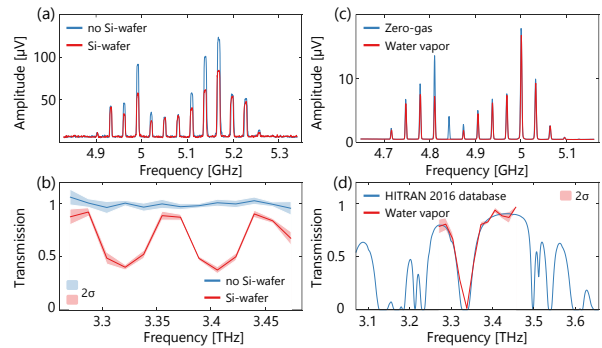

**Fig. 2.** (a) Beat note spectra acquired for 10 ms with (red) or without (blue) a tilted silicon wafer inserted in the beam path. The beat note amplitude attenuation is clearly observed. (b) Transmission spectra calculated from the beat note spectra in (a). A nearly flat transmission can be observed with a cleared beam path (blue) and the etalon structure originating from the 525 μm Si-wafer is shown in red. The shaded areas correspond to a 95% confidence interval. (c) Beat note spectra acquired for 20 ms with zero-gas (blue) or atmospheric water vapor at 23% relative humidity (red). (d) Transmission spectrum (red) together with a simulation based on parameters from the HITRAN [30]

2016 database (blue). The mode spacings were determined from Fig. 1(e).

The system was designed as shown in Fig. 1(a), where two chip-scale THz-QCL OFCs [see panel (b)] are arranged to emit anti-parallel THz beams that are collimated using a combination of silicon lenses and off-axis parabolic mirrors. The collimated THz light from one of the lasers, the signal OFC, is focused onto the sample. The radiation transmitted through the sample is re-collimated and combined with the light emitted from the local oscillator OFC using a silicon beam splitter. The dual-comb light is focused onto a sensitive hot electron bolometer used as a square-law mixer with a 6 dB bandwidth of 5 GHz. Due to the multi-heterodyne mixing process of the optical modes from both THz-QCL OFCs the spectral information from the THz frequency domain is imprinted in the radio-frequency (rf) modulation of the photocurrent, which can be conveniently digitized using an rf spectrum analyzer. The FTS spectra obtained for the two OFCs are shown in Fig. 1(c), where 13 and 16 modes can be observed for the two combs, respectively, resulting in an instantaneous optical coverage of 220 GHz for the DCS system. Figure 1(d) shows the interferograms for the spectra of Fig. 1(c). Figure 1(e), (f) show narrow and stable rf signals corresponding to the round-trip frequencies measured directly at the OFCs electrical terminals during operation, which, in combination with broadband FTS spectra, is indicative of comb operation.

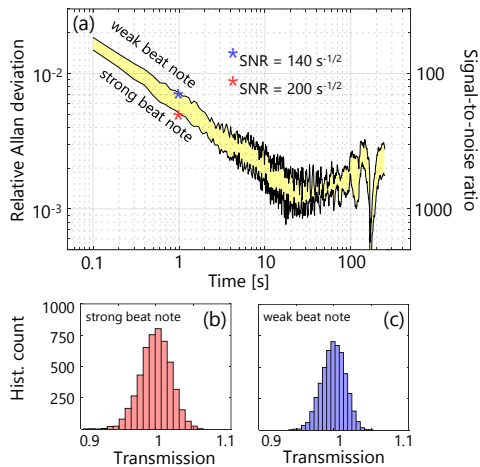

**Fig. 3.** (a) Allan deviation plot showing the relative stability of the beat note amplitudes as a function of integration time. The shaded yellow area represents the range of beat note SNRs of the system. (b), (c) Histograms of transmission values for the strong and weak beat notes of (a).

DCS spectra recorded for a duration of 10 ms are shown in Fig. 2(a), which indicates good signal-to-noise beat notes with an average contrast of more than 20 dB. To estimate the frequency resolution and visualize the behavior of the beat note spectra when the optical modes are affected by optical absorption, a tilted 525 μm silicon wafer is introduced into the beam path, which gives rise to a frequency-periodic etalon fringes. The attenuated beat notes are displayed in red. Figure 2(b) shows the corresponding transmission spectra, where the blue trace represents a clear beam path and the red trace the etalon transmission (95% confidence intervals are indicated by the shaded areas). Fig 2(c), (d) show the corresponding measurements for zero-gas and atmospheric water vapor (23% relative humidity). These results, together with the spectra of Fig. 1(c) and (e), are used for optical frequency calibration of the system.

To evaluate the short-term precision and long-term stability of the system we performed an Allan deviation analysis of the relative beat note amplitude stabilities was performed. Rf beat note spectra were recorded with a 100 ms time-resolution for a total acquisition time of 500 s. The results are shown in Fig. 3(a), where the shaded area represents the range of precisions observed for all beat notes. The strongest beat notes with SNRs (1σ) of ~200/√s result in proportionally better precisions as compared to the weaker ones with SNRs of ~140/√s. A drift appears after ~20 s of averaging, which is mainly attributed to mechanical vibrations induced by the cryopump. Figure 3(b), (c) show the histograms of transmission values used for the analysis in Fig. 3(a).

A sample of two common pharmaceutical excipients and an amino acid in the form of a pressed disc was prepared to demonstrate the hyperspectral imaging capabilities of the system. The disc comprised three zones with a 10% mass concentration of α-D-glucose monohydrate (GMH), α-D-lactose monohydrate (LMH), and L-histidine monohydrochloride monohydrate (LHHM) powder (Sigma-Aldrich, 99%) diluted in spectroscopic grade polyethylene and was pressed using 5 tons of pressure. The disc was placed in a sample holder with XYZ-translation capability, and the sample was raster-scanned across the focus of the THz beam with a step size of 0.5 mm resulting in a 81×53 pixel THz image acquired as a three-dimensional hyperspectral data cube using simultaneously acquired dual-comb spectra. The acquisition time was limited by the movement of the translation stage and the data transfer speed to approximately 0.3 s per pixel, resulting in a total acquisition time of 21 min for the hyperspectral image shown in Fig. 4(b). By upgrading the slow translation stage to faster raster-scan hardware similar images could potentially be obtained on the sub-second time-scale. Simple Gaussian smoothening and contrast enhancing sigmoid transformation was used to obtain the hyperspectral slices shown in Fig. 4(a). Images labeled R, G, and B show selected hyperspectral slices where the different spectral signatures of the material absorption shown in Fig. 4(c) were assigned distinct false colors. Two zones stand out due to distinct spectral trends in material absorption, whereas the third zone, as a result of a nearly feature-less absorption spectrum in this spectral region show less contrast. The RGB composite image of Fig. 4(b) is obtained by combining the R, G, and B log-transmission slices of Fig. 4(a), measured at 3.30 THz, 3.35 THz, and 3.45 THz, respectively. The three zones with different compounds are clearly identifiable in the composite image. The bottom-left zone with GMH appears as red in the transmission image due to a monotonically increasing absorption with frequency, as opposed to the bottom-right purple zone of LMH, which has the lowest attenuation in the blue (high-frequency) channel, and the strongest in green. Finally, the top LHHM zone appears as orange-white because it is the most transparent of all with slightly increased absorption in the blue channel. To demonstrate the resolution capabilities of the system, a reflective intensity image with ~12000 pixels of a US quarter was acquired [see Fig. 4(d)]. The inset shows a finer scan of the details of the feathers on the eagle's wing with <200 μm resolution.

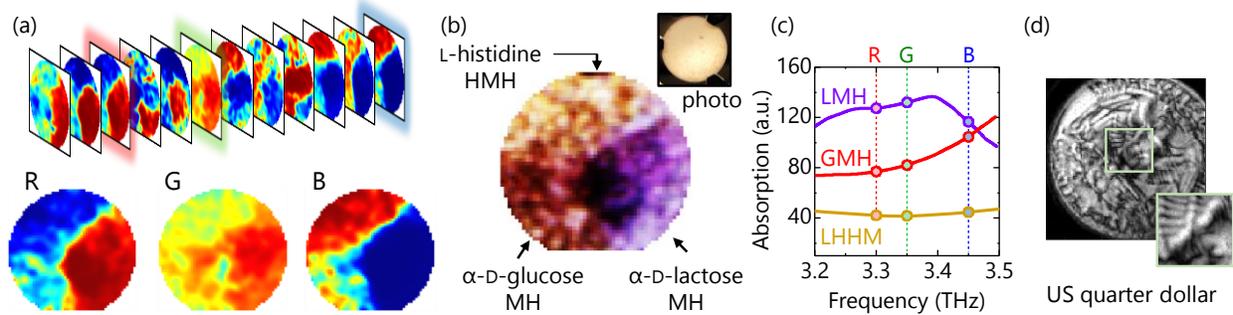

**Fig. 4.** (a) Individual mean-normalized absorbance hyperspectral slices corresponding to the beat notes shown in Fig. 2A. False colored R, G, and B show zoomed slices at three selected frequencies: 3.30 THz; 3.35 THz, and 3.45 THz respectively. The three zones are clearly identifiable. (b) False color RGB composite image composed from slices R, G, and B in transmission mode without mean removal. The inset shows a backlit photo of the pressed sample disc. (c) Reference spectroscopic data of α-D-glucose monohydrate (GMH) and α-D-lactose monohydrate (LMH) from Ref. [32], and L-histidine monohydrochloride monohydrate (LHHM) from Ref. [33] showing the frequency-dependent attenuation of the absorbers responsible for the different zone colors in (b). The vertical dashed lines and circles indicate the sampling points of the composite image channels. (d) 111×111-pixel reflective intensity imaging of a US quarter to demonstrate the resolution capabilities of the system at this wavelength. The inset shows a finer scan of the details of the eagle's wing.

In conclusion, we have experimentally demonstrated hyperspectral imaging system using dual chip-scale semiconductor laser frequency combs, all while operating in the challenging THz spectral region. An 81×53-pixel hyperspectral image was acquired through a raster scan of a solid pressed disc containing three differently absorbing compounds: α-D-glucose monohydrate, α-D-lactose monohydrate, and L-histidine monohydrochloride monohydrate. The THz sources operate at a center frequency of 3.4 THz and span approximately 220 GHz with an optical power of ~10 μW per mode, which is more than typically achieved with Fourier transform spectroscopy or time-domain spectroscopy at this wavelength. The acquisition speed of the instrument is currently limited by the slow mechanical raster scan, but assessments of the short-term precision of the system indicate that percentage level fractional absorption could be obtained on the sub-second time-scale with faster scanning. Future developments of the THz combs are expected to further increase the optical power and address the current limitations in optical bandwidth [33]. Already, octave-spanning multimode THz-QCLs have been demonstrated [34]. In addition, a reduction in the duty-cycle of the combs through pulsed operation can be used to increase the operating temperature, albeit with an accompanied loss in sensitivity. Even so, pulsed mode operation is likely the most viable path towards practical applications in bioimaging or the pharmaceutical quality control.

**Funding**

The work at Princeton was supported by the Defense Advanced Research Projects Agency (DARPA) SCOUT program (W31P4Q-16-1-0001), Thorlabs Inc. The work at MIT was supported by the Defense Advanced Research Projects Agency (DARPA), and the U.S. Army Aviation and Missile Research, Development, and Engineering Center (AMRDEC) through grant number W31P4Q-16-1-0001. The views and conclusions contained in this document are those of the authors and should not be interpreted as representing the official policies, either expressed or implied, of the Defense Advanced Research Projects Agency, the U.S. Army, or the U.S. Government. L.A.S. acknowledges support from the Kosciuszko Foundation Research Grant, and the Foundation for Polish Science within the START program. This work in this Report was performed, in part, at the Center for Integrated Nanotechnologies, an Office of Science User Facility operated for the U.S. Department of Energy (DOE) Office of Science. Sandia National Laboratories is a multimission laboratory managed and operated by National Technology and Engineering Solutions of Sandia, LLC., a wholly owned subsidiary of Honeywell International, Inc., for the U.S. Department of Energy's National Nuclear Security Administration under contract DE-NA-0003525.